\documentclass[trackchanges, twocolumn]{aastex701}

\usepackage[utf8]{inputenc}
\usepackage{natbib}
\usepackage{amssymb}
\usepackage{amsmath}
\usepackage{textgreek}

\makeatletter
\def\@submitted{Accepted to ApJ}
\makeatother

\begin{document}

\title{A Parsec-scale Circumstellar Shell Around VX Sgr: Mass Loss and RSG/TŻO Nature}

\author[orcid=0009-0008-7952-5702,sname='North America']{Julian Shapiro}
\affiliation{Department of Astronomy, University of California, Berkeley, CA 94720-3411, USA}

\email[show]{julianshapiro@berkeley.edu}  

\begin{abstract}

A thin, arc-shaped shell is discovered surrounding VX Sagittarii (VX Sgr), the largest known CSM shell around an RSG at a scale $r\sim0.59\,\mathrm{pc}$. The nebula has an extinction-corrected H$\alpha$ intensity $I_{\mathrm{H}\alpha}\sim38\,\mathrm{R}$ and an ionized mass $\sim0.1\,\mathrm{M_\odot}$. The slow $12\,\mathrm{km\,s}^{-1}$ peculiar velocity of VX Sgr and the lack of far-infrared emission in the arc indicate that it is not produced by a bow shock, but is likely instead steady RSG wind or ejecta that is photoionized by a nearby $\mathrm{O}$- or $\mathrm{B}$- type star. The nature of VX Sgr is debated in the literature as a red supergiant/hypergiant (RSG/RHG), super-AGB star, or Thorne--Żytkow Object (TŻO). Using the kinematic age of the Sgr OB1 association, the super-AGB model is disfavored. A spectroscopic test of the arc is proposed to potentially distinguish an RSG from a possible TŻO. The mass-loss rate is estimated from the geometry and emission measure of the shell, finding $\dot{M}\approx4\times10^{-5}\,\mathrm{M}_\odot\,\mathrm{yr}^{-1}$. The future interaction of the supernova (SN) ejecta of VX Sgr and the surrounding circumstellar material (CSM) is briefly explored through the mass ratio $\Lambda$ of the shell and SN ejecta; the interaction would likely not have a significant effect on the SN forward shock.
\end{abstract}



\section{Introduction}

VX Sagittarii (VX Sgr) is an extremely luminous, cool evolved star in our Galaxy. Over nearly a century of observations, VX Sgr has exhibited semi-regular variability, with active and quiescent phases. Its active photometric period is $\sim750$ days (e.g. \citealt{1971gcvs.book.....K, 10.1093/pasj/57.2.341, Tabernero_2021, Jadlovsky_2026}), dropping to $\sim320$ days while quiescent \citep{Jadlovsky_2026}. VX Sgr's temperature estimates range $\sim2900\mathrm{-}3700\,\mathrm{K}$ (e.g. \citealt{2007A&A...462..711G, 10.1093/mnras/stw3165, Tabernero_2021}), and it varies in spectral type from M5.5 at maximum to M9.8 at minimum \citep{1982MNRAS.198..385L}.

The distance to VX Sgr is historically disputed, ranging $0.3\mathrm{-2.0\,\mathrm{kpc}}$ in the literature \citep{Xi_Chen_2007}. The lower bound of photometric and parallax distances \citep{1974ApJ...188...75H, 2007A&A...474..653V} $\sim0.3\mathrm{-}0.8\,\mathrm{kpc}$ is likely contaminated by convection structures on the stellar surface (\citealt{Chiavassa_2011b, Chiavassa_2022}). VX Sgr is often considered to be a member of the Sgr OB1 association at $\sim1.7\,\mathrm{kpc}$ \citep{1972ApJ...172...75H}, though its grouping is not homogeneous and contains multiple kinematic subgroups \citep{1995AstL...21...10M, Tabernero_2021} that are likely at varying distances. The most reliable estimates emerge from recent studies of VX Sgr's $\mathrm{SiO}$ and $\mathrm{H_2O}$ masers. Using the statistical parallax of the $\mathrm{SiO}$ maser, \citet{Xi_Chen_2007} find $1.57\pm0.27\,\mathrm{kpc}$ and \citet{Su_2018} derive $1.10\pm0.11\,\mathrm{kpc}$. Through analysis of the $\mathrm{H_2O}$ maser proper motion, \citet{2003MNRAS.344....1M} determine $1.8\pm0.5\,\mathrm{kpc}$, \citet{1998ASPC..144..253M} find $1.7\pm0.3\,\mathrm{kpc}$, and \citet{Xu_2018} measure $1.56^{+0.11}_{-0.10}\,\mathrm{kpc}$. As a result of varying distance estimates, the radius and luminosity of VX Sgr are not well constrained in the literature, ranging $\sim1000\mathrm{-}2000\,R_\odot$ \citep{Chiavassa_2022} and $5.0\mathrm{-}5.6\,\log(\mathrm{L}/\mathrm{L}_\odot)$ \citep{Tabernero_2021}.

The nature of VX Sgr is debated due to its imprecise parameters, as well as its unusual spectral lines and variability. It is most often considered a red supergiant (RSG) or red hypergiant (RHG); when anchored by the $1.56^{+0.11}_{-0.10}\,\mathrm{kpc}$ maser distance by \citet{Xu_2018}, \citet{Jadlovsky_2026} argue that the corresponding luminosity implies a mass of $20\mathrm{-}40\,M_\odot$ \citep{Arroyo-Torres_2013, Arroyo-Torres_2015}, in excess of the typical AGB range. \citet{Tabernero_2021} argue that VX Sgr is a super-AGB star due to the presence of circumstellar \ion{Rb}{1} lines, unexpected for RSG/RHGs. The possibility of a Thorne--Żytkow Object (TŻO), a hypothetical star proposed by \citet{1977ApJ...212..832T}, is also considered. A TŻO would appear nearly identical to RSGs, only potentially distinguishable through spectral lines (e.g. \citealt{10.1093/mnras/263.4.817}), period \citep{10.1093/mnras/stad1977}, or multimessenger signals \citep{DeMarchi_2021}. \citet{Tabernero_2021} disfavor this scenario due to the lack of \ion{Li}{1}, predicted in some TŻO models (e.g. \citealt{10.1093/mnras/274.2.485}).

\begin{figure*}
    \centering
    \includegraphics[width=1\linewidth]{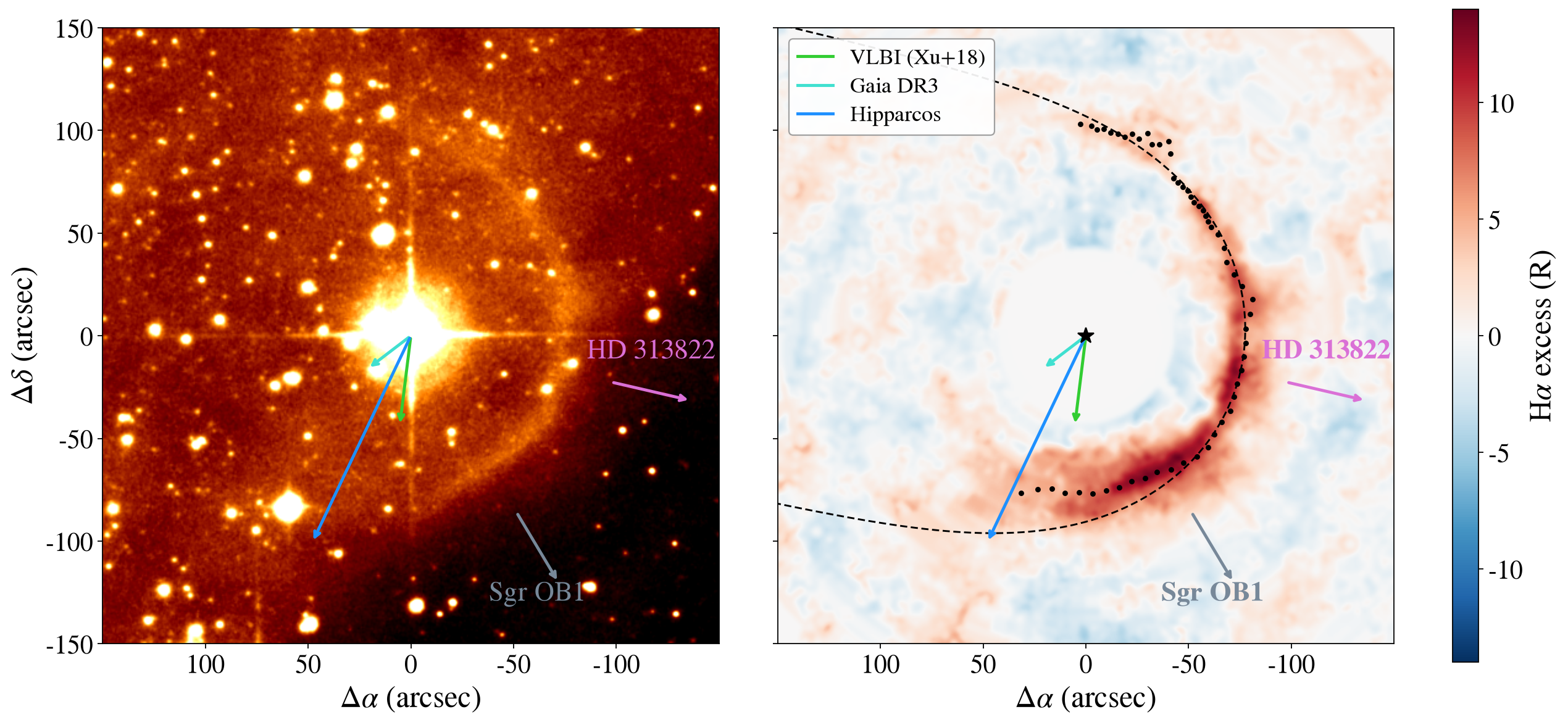}
    \caption{The arc-like H$\alpha$ nebula surrounding VX Sgr in the Super-COSMOS H-alpha Survey (left) and the star-removed H$\alpha$ excess mask in R (right). The pink arrow points towards the potential ionizing source HD 313822, and the grey arrow shows the direction to the center of Sgr OB1 from \citet{10.1093/mnras/staa454}. The proper motion vectors from VLBI (adopted; \citealt{Xu_2018}), Gaia DR3 \citep{Gaia_2016, 2023A&A...674A...1G}, and Hipparcos \citep{2007A&A...474..653V} are displayed and scaled to motion over $15\,\mathrm{kyr}$ for visibility. N is up and E is left. FOV = $5\times5$ arcmin.}
    \label{fig:vxsgr_image}
\end{figure*}

Circumstellar material is observed on the sub-arcsecond scale as compact nebulae around some RSG/RHGs, including VX Sgr, observed by \citet{2006AJ....131..603S} using the Hubble Space Telescope (HST). A handful of RSG/RHGs also have far larger structures; the runaway stars Betelgeuse \citep{1997AJ....114..837N}, $\mu$ Cep \citep{Cox_2012}, and IRC --10414 \citep{10.1093/mnras/stt1943} contain prominent bow shocks. Some RSGs with small peculiar velocities also demonstrate similar surrounding nebulae; notably, NML Cyg \citep{1982A&A...108..412H, 1983ApJ...267..179M} contains a shell detected in H$\alpha$ and $21\,\mathrm{cm}$ emission, potentially resulting from RSG mass-loss shaped by photodissociation from Cyg OB2. In this work, I report the discovery of arc-like H$\alpha$ emission surrounding VX Sgr. Similar to NML Cyg, VX Sgr is consistent with the motion of its association and the arc is thus most likely RSG wind or ejecta that is photoionized by hot $B$ or $O$ type stars, rather than shock ionized in the form of a bow shock.

RSG/RHGs undergo significant mass-loss prior to Type II-P core collapse, which shapes their subsequent path on the Hertzsprung--Russell diagram \citep{10.1093/mnras/stx3174}. The theoretical processes driving RSG winds are not well understood (see \citealt{2025Galax..13...72V} for a review on RSG mass loss), requiring stellar models to rely on empirical $\dot{M}$--luminosity relationships (e.g. \citealt{10.1093/mnras/staa255}) with only small, high-scatter observational samples (e.g. \citealt{1988A&AS...72..259D}). The H$\alpha$ shell surrounding VX Sgr provides a rare look into the mass-loss history of a likely RSG. The asymmetric arc may also play a role in shaping the following supernova remnant (SNR) ejecta.

\section{Data}

The environment of VX Sgr is within the coverage of multiple high-resolution H$\alpha$ surveys. The SuperCOSMOS H-alpha Survey (SHS; \citealt{2005MNRAS.362..689P, 10.1093/mnras/stt1986}) provides a $\sim1\,\mathrm{arcsec}$ resolution image of the arc from the $1.24\,\mathrm{m}$ UK Schmidt Telescope of the Anglo-Australian Observatory. SHS's R mag limit for point sources is $\sim20.5$. The region is captured by the $2.6\,\mathrm{m}$ VST/OmegaCAM Photometric H-alpha Survey (VPHAS+; \citealt{10.1093/mnras/stu394}), which offers a similar $\sim1\,\mathrm{arcsec}$ resolution alongside $u$, $g$, $r$, and $i$ band data. The target depth of VPHAS+ is $\gtrsim20\,\mathrm{mag}$ at $10\sigma$ in each filter. The VX Sgr CSM is also detected in the upcoming second data release of the MDW H-alpha Sky Survey (DR1, \citealt{Aftab_2026}; DR2, Aftab et al. in prep.).

\section{Ionized H$\alpha$ Arc}

\begin{figure*}
    \centering
    \includegraphics[width=1\linewidth]{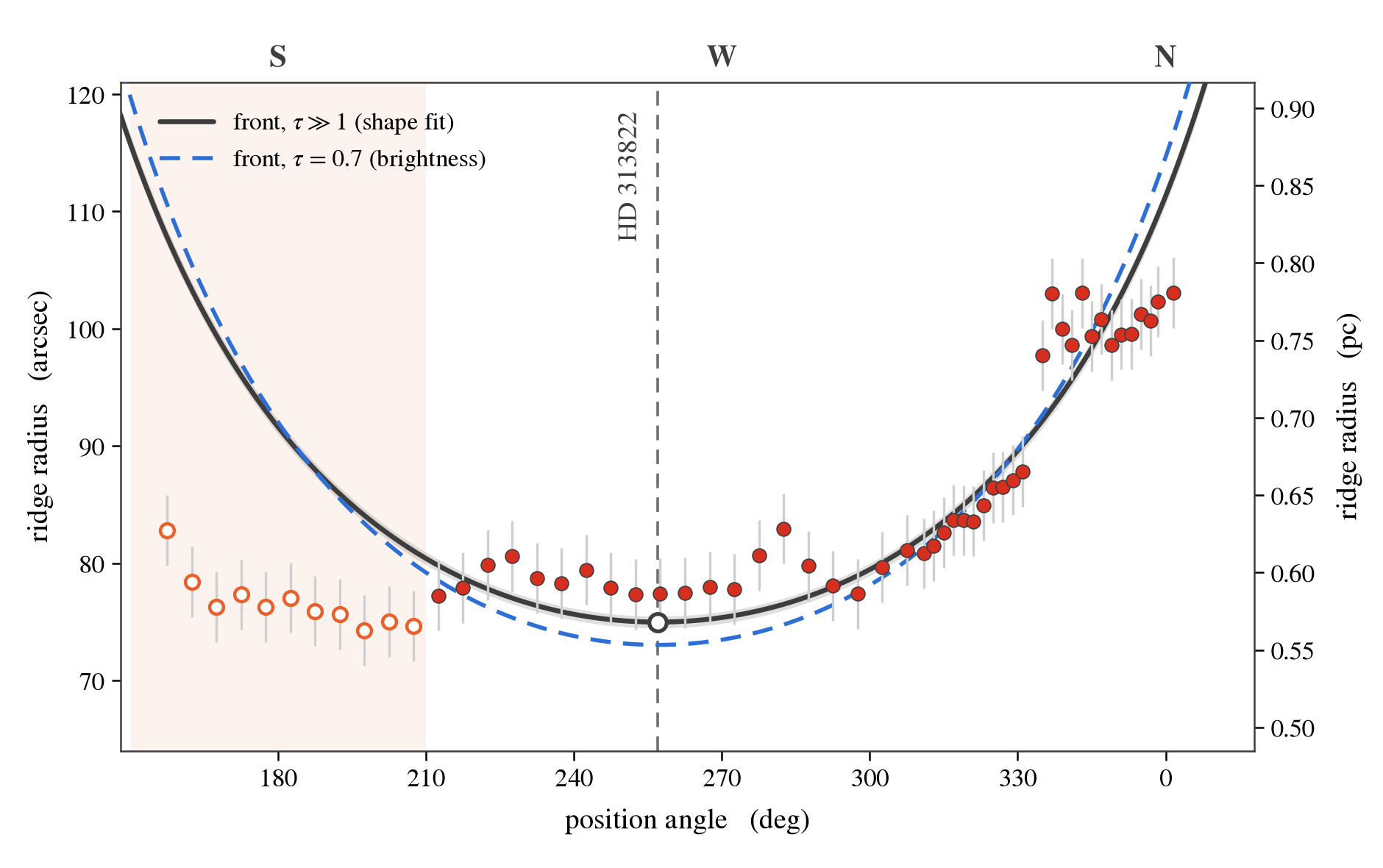}
    \caption{Profile of the position angle against arc ridge of detected bins in VPHAS+ with $1\sigma$ uncertainty. Orange points are highlighted $<210\degr$ where they deviate from the points $>210\degr$ in red. The arc shape \citep{1983ApJ...267..179M} defined by $\tau_\mathrm{shape}$ is shown in black, and the dashed blue line is determined from the measured $\tau_\mathrm{SB}$.}
    \label{fig:ridge_profile}
\end{figure*}
\subsection{Ionization Source}

Of the known red supergiant population, only a small sample contains extended large-scale structure. The prototypical RSG Betelgeuse contains a prominent arc detected in infrared, revealed to be a bow shock resulting from its $24\mathrm{-}33\,\mathrm{km\,s}^{-1}$ peculiar velocity relative to the ISM \citep{10.1093/pasj/60.sp2.S407,Cox_2012}. Observations from Herschel/PACS reveal its clumpy structure, multiple arcs, and linear bar \citep{Decin_2012}, some of which may be attributed to photoionization \citep{2014Natur.512..282M}. \citet{Cox_2012} conducted an extensive survey of detached shells around RSGs and AGB stars with Herschel\footnote{Herschel is an ESA space observatory with science instruments provided by European-led Principal Investigator consortia and with important participation from NASA.} \citep{Pilbratt_2010} PACS \citep{Poglitsch_2010}, revealing a similar bow shock around $\mu$ Cep, with a peculiar velocity of $\sim22\,\mathrm{km\,s}^{-1}$ (at $0.39\,\mathrm{kpc}$, $\sim30\,\mathrm{km\,s}^{-1}$ at $0.94\,\mathrm{kpc}$ from \citealt{10.1093/mnras/staa174}, shown in Figure \ref{fig:rsg_population}).

The first optical detection of an RSG bow shock in H$\alpha$ around IRC --10414 by \citet{10.1093/mnras/stt1943} raises the question of whether VX Sgr's arc is of the same nature. IRC --10414 is a clear runaway with an exceptionally high space velocity $\sim65\,\mathrm{km\,s}^{-1}$, approximately aligned with the orientation of the arc \citep{10.1093/mnras/stt1943}. Similar to distances, optical proper motion measurements of RSGs are subject to significant error. In the case of VX Sgr, the new Hipparcos reduction catalog estimates $3.20\pm3.08\,\mathrm{mas\,yr}^{-1}\,\alpha$ and $-6.71\pm1.72\,\mathrm{mas\,yr}^{-1}\,\delta$ \citep{2007A&A...474..653V}, while Gaia DR3 notes $1.40\pm0.20\,\mathrm{mas\,yr}^{-1}\,\alpha$ and $-1.07\pm0.15\,\mathrm{mas\,yr}^{-1}\,\delta$ \citep{Gaia_2016, 2023A&A...674A...1G}. Instead, the proper motion is adopted as $0.36\,\mathrm{mas\,yr}^{-1}$ pmE and $-2.92\,\mathrm{mas\,yr}^{-1}$ pmN from \citet{Xu_2018}, and the line of sight velocity is adopted as $5\,\mathrm{km\,s}^{-1}$ from \citet{2019ApJ...885..131R}. \citet{Xu_2018} use phase-referencing observations with the Very Long Baseline Array (VLBA) to measure the proper motion of a reference maser spot fitted to an expanding outflow model with the relative motions of 16 additional maser spots. The result provides a precise, outflow-corrected astrometric measurement of the proper motion of VX Sgr. As noted by the authors, the optical measurements (e.g. Hipparcos) of the system do not provide meaningful proper motion constraints, likely due to optical dust scattering and inhomogeneities in the CSM (e.g. \citealt{Zhang_2012b}). Using the corresponding $\mathrm{H_2O}$ maser VLBI distance of $1.56^{+0.11}_{-0.10}\,\mathrm{kpc}$ \citep{Xu_2018}, the peculiar velocity of VX Sgr is $\sim12\,\mathrm{km\,s}^{-1}$, in broad agreement with Sgr OB1 and inconsistent with a bow shock, especially considering its likely coincidence with the ionized \ion{H}{2} region LBN 38. An RSG bow shock requires the supersonic movement of the star through its medium. In an \ion{H}{2} region, the isothermal sound speed is high $\sim10\,\mathrm{km\,s}^{-1}$ \citep{Mackey_2015}, which is approximately consistent with the peculiar velocity of VX Sgr itself (Mach $\sim1$). RSG bow shocks are observed at Mach $\gtrsim3$ from their peculiar velocity measurements (e.g. \citealt{10.1093/mnras/stt1943}; see Section \ref{population}) even assuming a similarly warm medium. The peculiar velocity vector of VX Sgr is also at a position angle of $\sim179\degr$, inconsistent with the axis of the arc $\sim257\degr$, as shown in Figure \ref{fig:vxsgr_image}.

The VX Sgr shell is most likely photoionized in a case similar to NML Cygni. Both are also targeted in the Herschel/PACS survey \citep{Cox_2012}, but are reported as non-detections for IR shells. VLBA and VLA astrometry of NML Cyg finds a distance of $1.61^{+0.13}_{-0.11}\,\mathrm{kpc}$ and a low peculiar velocity $\sim8\,\mathrm{km\,s}^{-1}$ \citep{Zhang_2012}. The motion of NML Cyg is in agreement with the nearby Cyg OB2, with a separation of $\sim100\,\mathrm{pc}$, proposed to be the ionization source of its outer circumstellar nebula \citep{1983ApJ...267..179M, 2006AJ....131..603S}. At their adopted distances, NML Cyg and VX Sgr have similar luminosities $\log (L/\mathrm{L}_\odot)\sim5.4$ and $\sim5.3$, respectively. However, the circumstellar arc nebula of VX Sgr is $\sim2.1$ times larger in diameter and substantially fainter than the primary arc of NML Cyg, potentially representing an earlier RSG wind history or mass loss episode. NML Cyg's nebula contains further H$\alpha$ emission beyond its arc, which may be the result of ISM interaction or previous ejecta. The inner circumstellar envelope of VX Sgr, as imaged using HST by \citet{2006AJ....131..603S}, is a nearly spheroidal envelope, whereas NML Cyg contains an asymmetric nebula consistent with the orientation of its $\mathrm{H_2O}$ masers and the larger \ion{H}{2} region.

Potential photoionization sources are searched for in SIMBAD \citep{2000A&AS..143....9W} as hot stars within the Sgr OB1 association, limited by Gaia DR3 \citep{Gaia_2016,2023A&A...674A...1G} $\varpi=0.45\mathrm{-}0.95\,\mathrm{mas}$. The spectroscopically classified $\mathrm{OB}$ star HD 313822 is identified as a strong candidate, with a Gaia DR3 parallax distance $\sim1.5\,\mathrm{kpc}$ and aligned with the arc orientation, shown in Figure \ref{fig:vxsgr_image}. The Gaia Extended Stellar Parameterizer for hot stars (ESP-HS; \citealt{Creevey_2023}) estimates an effective temperature of $35810\pm438\,\mathrm{K}$, corresponding to a spectral class $\mathrm{O7-O7.5}\,\mathrm{III}$, for which the integrated Lyman-continuum photon rate is $\log \mathrm{Q}_0\approx49.0\mathrm{-}49.1\,\mathrm{s}^{-1}$ \citep{Martins_2005}. The total ionizing flux at VX Sgr is computed as $I=\mathrm{Q}/4\pi \mathrm{D}^2$, estimating $\sim6\times10^{8}\,\mathrm{ph\,cm^{-2}\,s^{-1}}$. The distances of $1.46\pm0.04\,\mathrm{kpc}$ to HD 313822 (Gaia DR3; \citealt{2021AJ....161..147B}) and $1.56^{+0.11}_{-0.10}\,\mathrm{kpc}$ to VX Sgr \citep{Xu_2018} are consistent within their uncertainties, and are not precise enough to determine a separation. The transverse distance between the two stars is $\gtrsim12.4\,\mathrm{pc}$ from their angular separation. Considering the observed arc shape, it is assumed that the inclination angle of the stars is low; if one star were at a substantially greater line of sight distance, the projected surface would appear nearly circular rather than asymmetric \citep{2006AJ....131..603S}. Thus, $\mathrm{D}\sim12.4\,\mathrm{pc}$ is adopted, which should be considered a lower limit.

\begin{figure*}
    \centering
    \includegraphics[width=1\linewidth]{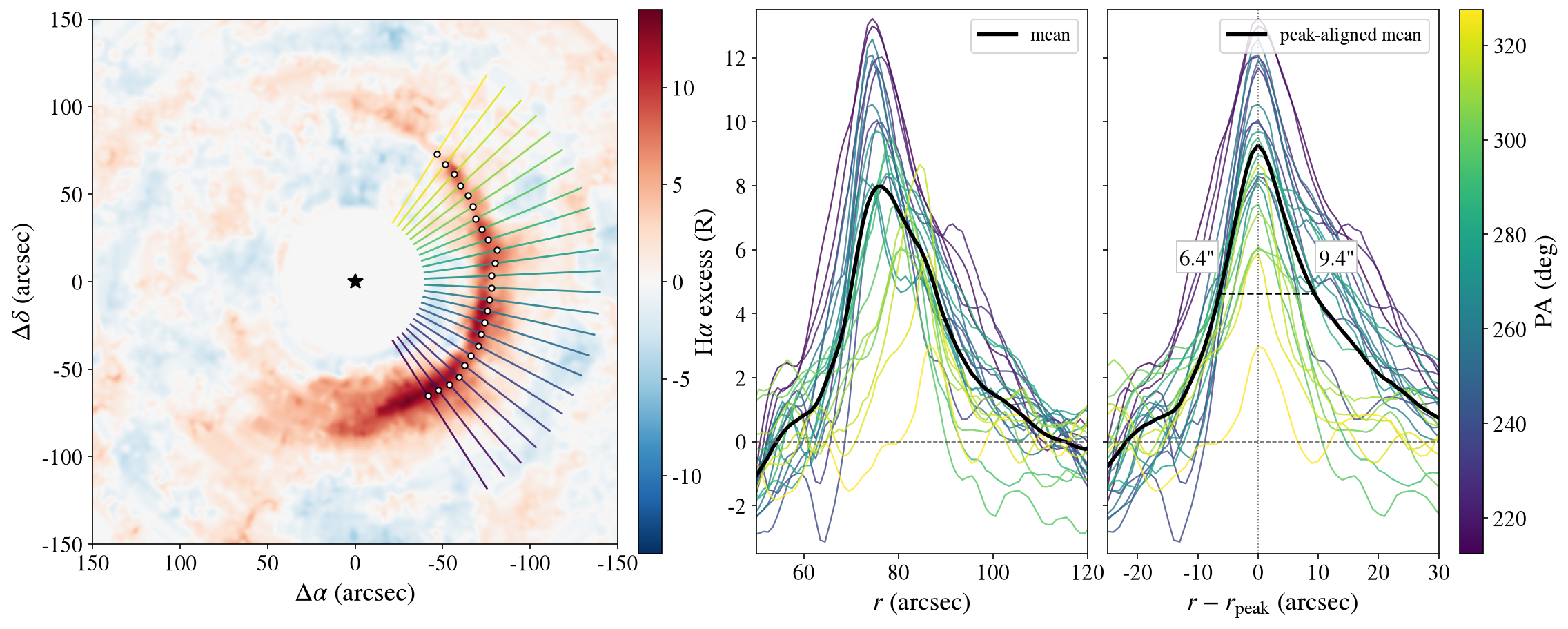}
    \caption{The H$\alpha$ excess mask from Figure \ref{fig:vxsgr_image} with $5\degr$ radial profiles for the PA range $210\mathrm{-}330\degr$ (left). The first plot on the right displays the H$\alpha$ intensity along the radial profiles, colorized by PA, and their mean. The second plot shows the profiles aligned to their peak values and the measured half width at half maxima.}
    \label{fig:profiles}
\end{figure*}

\subsection{Arc Properties}

To isolate the H$\alpha$ arc, stellar point sources are identified using a median filter ($9''$) and sigma clipping in the VPHAS+ cutout images. Any residuals above $3\sigma$ and larger than $60\,\mathrm{px}$ are removed, and the image is smoothed by $2\,\mathrm{px}$. $30\degr$-wide radial profiles are produced with $15\degr$ overlap. A parabola is fitted to each profile using ordinary least squares, and the excess map is obtained as the difference between the observed profile and parabola, shown in the right panel of Figure \ref{fig:vxsgr_image}.

Though VPHAS+ is of a higher resolution than SHS, it is not calibrated to Rayleighs (R) and is affected by artifacts from VX Sgr. Thus, the SHS survey is used for the intensity measurement by placing the image on a R scale through regression against continuum-subtracted SHASSA data \citep{Gaustad_2001}. The SHS H$\alpha$ intensity of the CSM is $I_{\mathrm{H}\alpha}\sim25^{+6}_{-5}\,\mathrm{R}$ within the VPHAS+-defined mask, and $\sim32^{+8}_{-6}\,\mathrm{R}$ at the arc apex.

The $7\,\mathrm{nm}$ bandpass of the SHS data, centered at $659\,\mathrm{nm}$, contains emission from the [\ion{N}{2}] $654.8$, $658.4\,\mathrm{nm}$ lines, which are contaminants of the H$\alpha$ intensity. The filter transmits $82\%$ and $50\%$ at the two [\ion{N}{2}] lines, respectively. Observations of ionized CSM shells around RSGs have detected high ratios of [\ion{N}{2}]/H$\alpha$, potentially due to nitrogen enrichment from the RSG dredge-up \citep{10.1093/mnras/stt1943}. The spectroscopic ratio [\ion{N}{2}] $654.8+658.4\,\mathrm{nm}$/H$\alpha\,656.3\,\mathrm{nm}$ of IRC --10414 is $\sim2.42$ \citep{10.1093/mnras/stt1943}, and of W26 is $\sim2.94$ \citep{2014MNRAS.437L...1W}. A mean $2.68\pm0.26$ is adopted as an estimate for VX Sgr. The calibration is applied as Equation 2 from \citet{10.1093/mnras/stt1986}, and multiplied by $1.13$ due to the [\ion{N}{2}]/H$\alpha\sim0.4$ in the SHASSA background, providing $I_{\mathrm{H}\alpha}\sim9.9^{+2.4}_{-2.4}\,\mathrm{R}$ and $I_{\mathrm{H}\alpha}\sim12.3^{+3.0}_{-2.9}\,\mathrm{R}$ for the full mask and apex, respectively.

Using the estimated visual extinction of VX Sgr $\sim1.5\,\mathrm{mag}$ \citep{1982MNRAS.198..385L}, these values are corrected to $\sim31^{+17}_{-11}\,\mathrm{R}$ and $\sim38^{+21}_{-13}\,\mathrm{R}$. The emission measures are $\mathrm{EM}\approx2.75\,I_{\mathrm{H}\alpha}=84^{+46}_{-30}\,\mathrm{pc\,cm}^{-6}$ and $105^{+57}_{-37}\,\mathrm{pc\,cm}^{-6}$ assuming $T_e\approx10^4\,\mathrm{K}$.

The CSM follows a nearly uniform rounded geometry along its ionization surface. It is compared to a cometary arc in Figure \ref{fig:ridge_profile}. The points $\gtrsim210\degr$ are consistent with the arc; though, the highlighted fainter region of CSM $\lesssim210\degr$ has points of lower radii, potentially due to differences in the densities of the CSM material and the surrounding medium. The fit is tested with a $\chi^2$ test using two free parameters: the arc radius $r_0$ and a shape parameter $\tau_\mathrm{shape}=\alpha n_0 r_0/3v$ \citep{1983ApJ...267..179M}; for a shape fit to the observed ridge at regions $>210\degr$, $\chi^2$ decreases with $\tau$, and the shape thus provides a lower limit $\tau_\mathrm{shape}\gtrsim20$, and $\chi^2/\mathrm{df}=1.70$.  If $\tau$ is instead estimated using $n_0$ (described in Section \ref{mass_loss}), $r_0=0.59\,\mathrm{pc}$, and $v=24.3\,\mathrm{km\,s}^{-1}$, the shape is computed as $\tau_\mathrm{SB}=0.7^{+0.2}_{-0.1}$. This is discrepant with the fit using two free parameters, which may arise due to a slower wind speed at the front than the $\mathrm{CO}$ velocity or a dense shell rather than the basic $r^{-2}$ wind model assumed in this work. However, as shown in Figure \ref{fig:ridge_profile}, the discrepancy in $\tau$ has little effect on the front shape, as the two curves are broadly consistent.

The measured properties of the VX Sgr arc are provided in Table \ref{tab:properties}.

In the photoionization front model of \citet{1983ApJ...267..179M}, it is expected that the H$\alpha$ intensity would drop off relatively sharply upstream (on the inside of the front) from the $r^{-2}$ wind and more smoothly on the outside of the shell. For a bow shock, the front is instead expected to be compressed downstream (e.g. \citealt{Wilkin_1996}). This is tested quantitatively on the VX Sgr arc using $5\degr$ profiles of the H$\alpha$ excess in the PA range $210\mathrm{-}330\degr$, shown in Figure \ref{fig:profiles}. The mean of the radial profiles aligned by their peaks, displayed in the right plot, demonstrates a sharper dropoff upstream than downstream, consistent with external photoionization. The half width at half maxima (HWHM) are ${6.4^{+0.3}_{-0.2}}''$ and ${9.4^{+1.0}_{-0.9}}''$ for the inside and outside ends from the mean peak, respectively, with errors estimated from bootstrap resampling of the profiles.

\begin{deluxetable}{lcc}
\setlength{\tabcolsep}{18pt}
\tabletypesize{\scriptsize}
\tablecaption{VX Sgr Arc Properties\label{tab:properties}}
\tablehead{
  \colhead{Parameter} &
  \colhead{Units} &
  \colhead{Value}
}
\startdata
$I_{\rm H\alpha,\,\rm{rim}}$ & R & $9.9^{+2.4}_{-2.4}$\\
$I_{\rm H\alpha,\,\rm{apex}}$ & R & $12.3^{+3.0}_{-2.9}$\\
$I_{\rm H\alpha,\,\rm{rim,\,A-corr.}}$ & R & $31^{+17}_{-11}$\\
$I_{\rm H\alpha,\,\rm{apex,\,A-corr.}}$ & R & $38^{+21}_{-13}$\\
$A_V$                    & mag                  & $1.5$ \\
$r_0$ & arcsec & $77.5\pm0.3$\\
$r_0$  & pc & $0.59\pm0.04$\\
$\tau_\mathrm{shape}$ & --- & $\gtrsim20$\\
$\tau_\mathrm{SB}$ & --- & $0.7^{+0.2}_{-0.1}$\\
Angular extent & deg & $204$\\
Symmetry axis & deg & $\sim257$\\
$\chi^2/\mathrm{df}$ (\textgreater 210\degr) & --- & 1.70\\
$\chi^2/\mathrm{df}$ (all) & --- & 6.29
\enddata
\tablecomments{Distance-dependent properties are measured adopting $1.56^{+0.11}_{-0.10}\,\mathrm{kpc}$ \citep{Xu_2018}.}
\end{deluxetable}

\begin{figure*}
    \centering
    \includegraphics[width=1\linewidth]{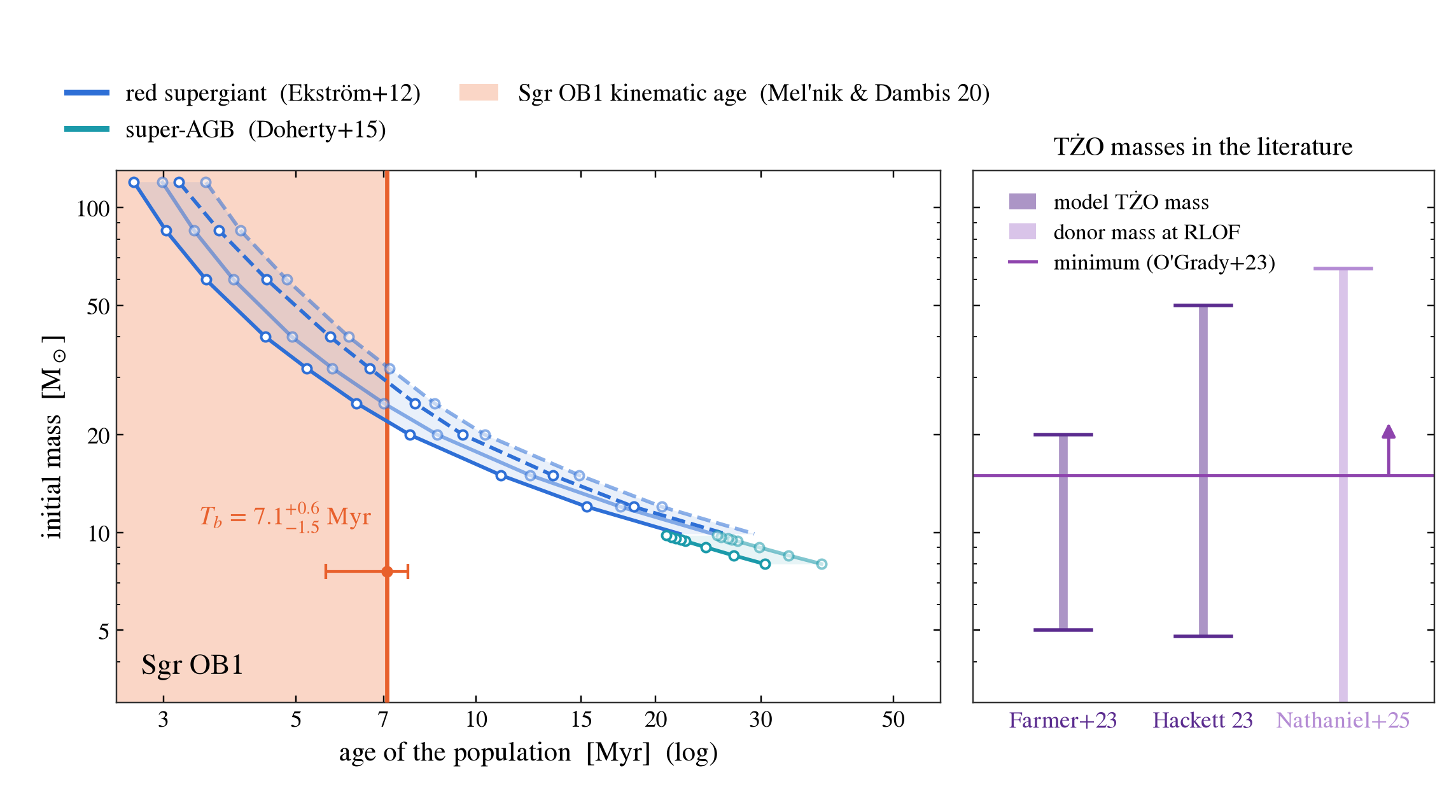}
    \caption{Possible initial masses of red supergiants (blue) from \citet{Ekstrom_2012} models $\geq9.9\,\mathrm{M_\odot}$ and AGB stars from \citet{10.1093/mnras/stu2180} models within $8.0\mathrm{-}9.8\,\mathrm{M_\odot}$ (teal) by age (left panel). The solid blue lines are non-rotating models, and the dashed blue lines are rotating. The darker blue and teal lines represent the end of core $\mathrm{H}$-burning; the lighter blue line displays the end of core $\mathrm{He}$-burning, and the lighter teal line shows the total lifetime. The right panel displays TŻO models from \citet{10.1093/mnras/stad1977, hackett_2023, Nathaniel_2025} over the same initial mass range. The age estimate of the Sgr OB1 association from \citet{10.1093/mnras/staa454} is highlighted in orange on the left, consistent only with RSGs and potentially TŻOs.} \label{rsg_agb_tzo}
\end{figure*}

\section{RSG, AGB, or T\.{Z}O?}

\subsection{On the Super-AGB Model}

The nature of VX Sgr has been debated in numerous studies, largely due to its poorly constrained distance. Historically, it is most often assumed to be a red supergiant or hypergiant due to its implied association with Sgr OB1 (e.g. \citealt{1972ApJ...172...75H}), assumed to be at $\sim1.7\,\mathrm{kpc}$, requiring a greater luminosity than the AGB limit $\log (L/\mathrm{L}_\odot)\approx5.1$ \citep{Groenewegen_2009}. Subsequent $\mathrm{SiO}$ and $\mathrm{H_2O}$ maser distance measurements $\sim1.6\mathrm{-}1.8\,\mathrm{kpc}$ (e.g. \citealt{Xi_Chen_2007}) supported this. However, later disagreement between more recent measurements, notably $1.10\pm0.11\,\mathrm{kpc}$ \citep{Su_2018} and $1.56^{+0.11}_{-0.10}\,\mathrm{kpc}$ \citep{Xu_2018}, has opened discussion as to whether VX Sgr may be a super-AGB star, which may be marginally permitted at closer distances.

VX Sgr exhibits unusual Mira-like variability and atmospheric properties, as noted by \citet{1972ApJ...172L..59H,1982MNRAS.198..385L, Chiavassa_2010}, though the authors still support an RSG nature. \citet{Tabernero_2021} argues that its Mira-like variability, supported by its line-doubling during its active phases, alongside the enhancement of \ion{Rb}{1}, assumed to be from $s$-process elements, supports an extremely high mass $\gtrsim10\,\mathrm{M}_\odot$ super-AGB star. However, this interpretation relies on a literature-weighted average distance $\sim1.44\,\mathrm {kpc}$, which is shifted down due to the \citet{Su_2018} $1.10\pm0.11\,\mathrm{kpc}$ measurement that is potentially an outlier. VX Sgr is most likely a member of the association at $\gtrsim1.3\,\mathrm{kpc}$, as it is not a runaway and has a low peculiar velocity $\sim12\,\mathrm{km\,s}^{-1}$. \citet{Arroyo-Torres_2013, Arroyo-Torres_2015} places VX Sgr in the HR diagram using estimates of its fundamental parameters to estimate an initial (birth) mass $\sim20\mathrm{-}40\,\mathrm{M}_\odot$.

\citet{Tabernero_2021} also assesses VX Sgr's candidacy as a Thorne--Żytkow Object (TŻO; \citealt{1975ApJ...199L..19T, 1977ApJ...212..832T}). These objects are theorized to form as a binary merger product of RSG cores and neutron stars, though their formation routes have been debated (e.g. see \citealt{OGRADY2026336}). \citet{1975ApJ...199L..19T} explored the possibility of a neutron star inspiraling into a stellar envelope, producing a TŻO during the mass transfer event if the common-envelope is not ejected. Alternatively, collisions can lead to the formation of a close binary and common envelope, similarly resulting in a potential TŻO (e.g. \citealt{10.1093/mnras/172.1.15P,ray1987evolution}). Under both formation routes, the resulting object would be very difficult to observationally distinguish from typical RSGs, though may contain $irp$ (interrupted rapid proton) process chemical enhancement based on the model. The lifespans of TŻOs across formation models are $\sim10^5\mathrm-10^6\,\mathrm{yr}$, and they would undergo similar mass-loss events to RSGs \citep{OGRADY2026336}.

The age of the stellar association of VX Sgr provides an estimate of its mass independent of \citet{Arroyo-Torres_2013, Arroyo-Torres_2015}. \citet{10.1093/mnras/staa454} studies the internal motions of Sgr OB1 to estimate its kinematic age using two methods--$T_b$, using the association's expansion parameters, and $T_\star$, with backtracking stellar positions--finding $T_b=7.1^{+0.6}_{-1.5}\,\mathrm{Myr}$ and $T_\star < 1\,\mathrm{Myr}$. \citet{10.1093/mnras/285.3.479} previously tested these methods, finding that the expansion method overestimates the kinematic age, while backtracking underestimates it. Thus, the two methods can be considered an approximate suitable age bracket for Sgr OB1.

For the approximately solar metallicity ($Z\sim0.02$) of VX Sgr \citet{Tabernero_2021}, the mass range of a super-AGB star is $\sim8.0\mathrm{-}9.8\,\mathrm{M_\odot}$ \citep{10.1093/mnras/stu2180}. \citet{Ekstrom_2012} computes grids of stellar evolutionary tracks, rotating and non-rotating, with a wide range $0.8\mathrm{-}120\,\mathrm{M_\odot}$ of mass inputs. For a super-AGB star, the age is considered to be the sum of the models at the end of H-burning and the end of He-burning. Interpolating the super-AGB mass range from the $7,\,9,$ and $12\,\mathrm{M_\odot}$ models, the total age requirement is $25.4\,\mathrm{-}37.5\,\mathrm{Myr}$ for a non-rotating star and $29.8\,\mathrm{-44.3}\,\mathrm{Myr}$ for a rotating one. The youngest possible super-AGB star far exceeds the upper limit of the age of the Sgr OB1 association and is thus not feasible, shown in Figure \ref{rsg_agb_tzo}, and in agreement with \citet{Arroyo-Torres_2013, Arroyo-Torres_2015}. It also does not permit an extraordinarily high-mass super-AGB star of $10\mathrm{-}12\,\mathrm{M}_\odot$ proposed by \citet{Tabernero_2021}.

\subsection{RSG/RHG or Thorne--Żytkow Object?}

The kinematics permit a mass $\gtrsim20\,\mathrm{M_\odot}$ with no defined upper limit, which is compatible with the range of high-mass RSGs or RHGs. However, it also cannot exclude the possibility of a TŻO due to the wide range of masses predicted by models (e.g. \citealt{10.1093/mnras/stad1977, Nathaniel_2025}), which often overlap with those of typical RSG/RHGs.

Theoretical models often propose high chemical abundances of lithium and $irp$-process elements in TŻOs as a potential observational signature (e.g. \citealt{1994ApJ...420..364B, 10.1093/mnras/274.2.485}). \citet{doi:10.1126/science.1133706, Tabernero_2021} finds that VX Sgr contains prominent \ion{Rb}{1}, which is not detected in any other RSG/RHG. \citet{Tabernero_2021} attributes this to $s$-process enrichment from a third dredge-up in an AGB. However, as an AGB star is not compatible with VX Sgr's mass, the \ion{Rb}{1} detection remains poorly explained. As Rb is an $irp$-process element, it is predicted to be enhanced in TŻO atmospheres (e.g. \citealt{1994ApJ...420..364B}) and was among the lines used to propose HV 2112 as a TŻO candidate \citep{10.1093/mnrasl/slu080}. It is important to note, though, that the \ion{Rb}{1} absorption is primarily circumstellar \citep{Tabernero_2021}, and He-rich models from \citet{10.1093/mnras/stad1977} do not necessarily produce Rb.

Recent ALMA $213\mathrm{-}270\,\mathrm{GHz}$ observations from the ATOMIUM program \citep{Wallstrom_2024} have unveiled the $\mathrm{CO}$, $\mathrm{SiO}$, $\mathrm{AlO}$, $\mathrm{AlOH}$, $\mathrm{AlF}$, $\mathrm{H}_2\mathrm{S}$, $\mathrm{PO}$, $\mathrm{OH}$, $\mathrm{TiO}$, $\mathrm{TiO}_2$, $\mathrm{HCN}$, $\mathrm{SO}$, $\mathrm{SO}_2$, $\mathrm{SiS}$, $\mathrm{CS}$, and $\mathrm{H}_2\mathrm{O}$ content within the observed bandpass. Within the sample of both RSG and AGB stars, VX Sgr has an anomalously high ratio of $^{29}\mathrm{Si}/^{30}\mathrm{Si}=4\pm1$, compared to the range $1\mathrm{-}2$ for typical oxygen-rich and S-type AGB stars. Potentially more useful for assessing the object as a TŻO candidate is the low ratio of $^{28}\mathrm{Si}$ to $^{29}\mathrm{Si}$ and $^{30}\mathrm{Si}$, as proposed by \citet{1995A&A...303L..25V}. It will be important to model these relative abundances in more detail, which is beyond the scope of this paper. However, the low ratio can also be mimicked by quenching and boosting mechanisms associated with line overlaps \citep{Lewis_2026}. Notably, $\mathrm{TiO}_2$ is also detected in the ALMA observations of VX Sgr. \citet{10.1093/mnras/stad1977} proposes that an abundance of $\mathrm{TiO}_2$ is expected to form in their TŻO models, and that $^{44}\mathrm{TiO}_2$ may be an observational signal of TŻOs. Unfortunately, $^{44}\mathrm{TiO}_2$ is beyond the detection capabilities of ALMA; $\mathrm{TiO}_2$ is also produced in RSGs, e.g., VY CMa \citet{Kaminski_2013} and thus is consistent with both.

The pulsation of VX Sgr, demonstrating both active Mira-like variability periods and quiescent phases, is unusual by the standards of RSGs. It may demonstrate some parallels to WOH G64 in the Large Magellanic Cloud, which previously exhibited similar Mira-like behavior as an RSG, with an $\sim850$ day period and shifts in magnitude $\Delta V\sim2\,\mathrm{mag}$, up until 2014 \citep{2026NatAs..10..702M}. It has since transitioned into more irregular variability and weakened pulsations (e.g. \citealt{2024A&A...691L..15O, 2026MNRAS.546ag012V, 2026NatAs..10..702M}).

Some TŻO models predict irregular, large-amplitude surface pulsations \citep{1975ApJ...199L..19T, hackett_2023, 10.1093/mnras/stad1977, Nathaniel_2025}. The hypothetical star is generally considered to be stable \citep{1975ApJ...199L..19T, 1977ApJ...212..832T, 1994ApJ...420..364B}, and any pulsations of TŻOs driven by the opacity feature in the hydrogen ionization zone of the envelope would be RSG-like \citep{hackett_2023} and a challenge to differentiate.

If the circumstellar shell of VX Sgr was produced through RSG wind or ejections, then it provides a unique testing ground for the TŻO hypothesis. The convection of $irp$-process products extends from the nuclear burning region to the stellar surface \citep{1994ApJ...420..364B}, producing the detectable $rp$-isotopes in the envelope \citep{1995A&A...303L..25V}. Following mass-loss, this enrichment would remain in the subsequent CSM shell. This parallels studies of $s$-process enrichment of planetary nebulae, where high resolution ($R>20000$) nebular spectroscopy is used to study the chemical properties of AGB progenitors (e.g. see \citealt{2020Galax...8...50S} for a review). \citet{10.1093/mnras/263.4.817} and \citet{1994ApJ...420..364B} predict enhancements relative to solar from $irp$-process $[\mathrm{Kr}/\mathrm{H}]=+4.2$ and $+2.0$, $[\mathrm{S}/\mathrm{H}]=+0.8$ and $+0.1$,  $[\mathrm{Cl}/\mathrm{H]=+2.0}$ and $+1.1$, and $[\mathrm{Ar}/\mathrm{H}]=+1.4$ and $+0.3$, respectively. Optical studies by \citet{1994A&A...283..593P, 2001ApJ...550L.223D, 2007ApJ...659.1265S, Sterling_2007, Sterling_2015, Mashburn_2016} demonstrate the methodology to distinguish enhancements at this scale with $\mathrm{Kr}$, $\mathrm{Xe}$, $\mathrm{Br}$, and $\mathrm{Se}$ in planetary nebulae and \ion{H}{2} regions. 

\citet{10.1093/mnras/stad1977} conducted the strongest existing test of VX Sgr as a TŻO candidate, finding consistency with their most $\mathrm{He}$-rich model, assuming its uncertain $\sim28000$ day pulsation period described by \citet{Tabernero_2021}. The authors detect the signal of this potential period at an amplitude of $\sim1\,\mathrm{mag}$ from the AAVSO light curve and demonstrate consistency with a sinusoidal toy model. However, they note that the feature cannot yet be confirmed, as the period is comparable to the time VX Sgr has been observed and thus could be an artifact. If it is a true period, the $\sim28000$ day period would need to have increased with the age of the TŻO from typical $\sim1000\mathrm {-}2000$ day fundamental pulsations, which is possible from high $\mathrm{He}$ production in the companion \citep{10.1093/mnras/stad1977}. This may be reflected in the $[\mathrm{He}/\mathrm{H}]$ abundance in the CSM of VX Sgr. 

Another potentially strong route for follow-up is a targeted neutrino observatory pointing with instruments such as Super-Kamiokande and IceCube, proposed by \citet{Martínez-Miravé_2025}. The hypercritical accretion rates onto a neutron star core $>10^{-4}M_\odot\,\mathrm{yr}^{-1}$ are 3 to 13 orders of magnitude greater than the Eddington rate, resulting in a photon trapping radius far larger than that of the NS, and thus the accretion energy being released as all flavors of neutrinos $1\,\mathrm{-}100\,\mathrm{MeV}$ \citep{1991ApJ...376..234H, Martínez-Miravé_2025}. Given the relatively close proximity to VX Sgr, a dedicated search should be able to identify the source as a TŻO if the accretion rate is $>10^{-2}\,\mathrm{M_\odot yr}^{-1}$ \citep{Martínez-Miravé_2025}.


\begin{figure}
    \centering
    \includegraphics[width=1\linewidth]{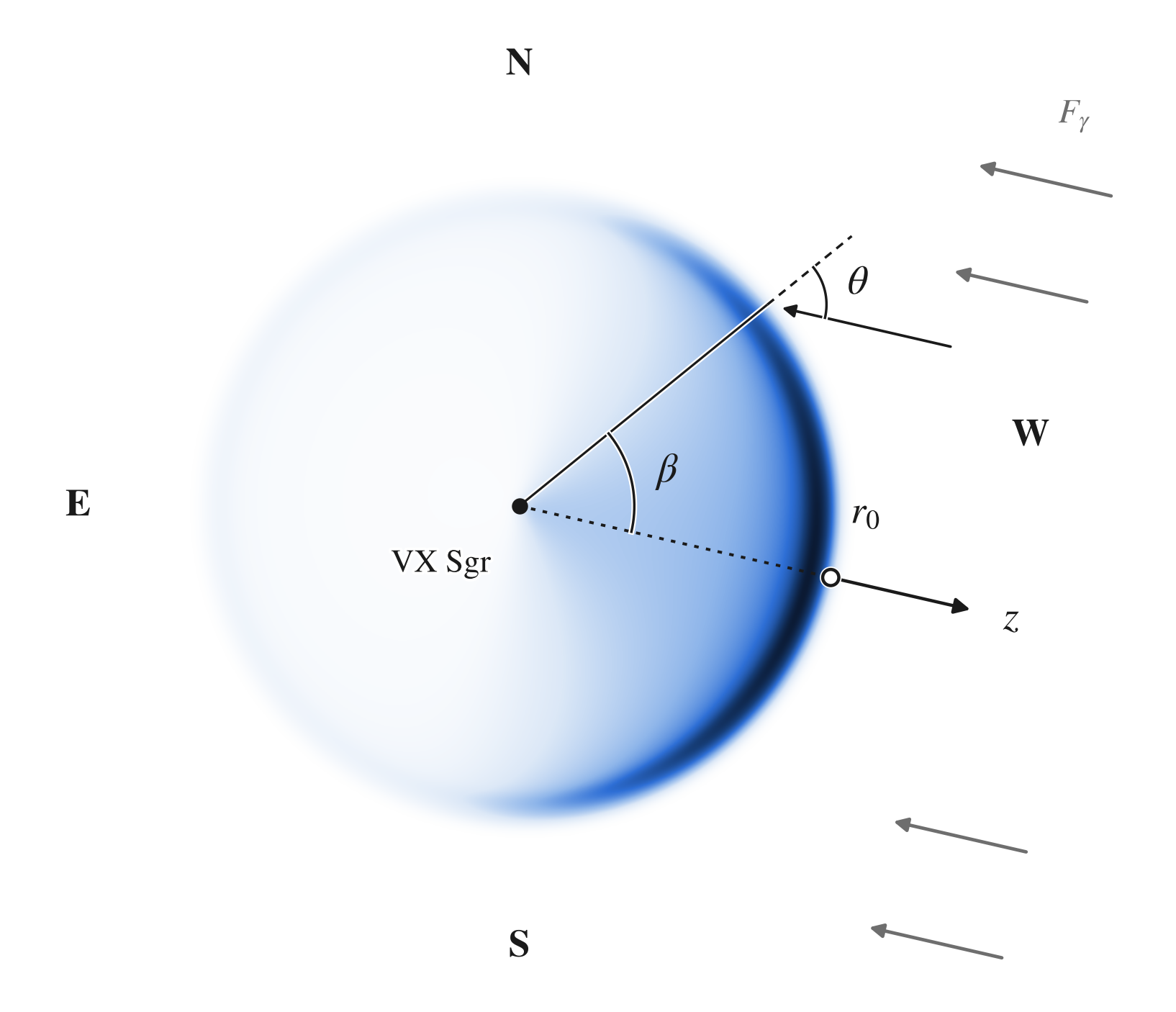}
    \caption{Illustration of the VX Sgr CSM nebula and the geometric parameters $\beta,\,\theta,\,r_0,$ and $z$ from \citet{1983ApJ...267..179M}. The figure is stylistically inspired by \citet{Orlando_2022}.}
    \label{fig:geometry}
\end{figure}

\section{Mass-loss}
\label{mass_loss}

The geometric properties of the photoionization arc surrounding VX Sgr can be used to probe the mass-loss rate. The incident ionizing flux $I$ in photons $\mathrm{cm}^{-2}\,\mathrm{s}^{-1}$ is defined by \citet{1983ApJ...267..179M} as
\begin{equation}
\begin{split}
I=\alpha\left(\frac{\dot{N}_w}{4\pi v}
\right)^2
\int^\infty_0 \left(r^2+z^2+2rz\cos \beta
\right)^{-2}dz\\
+\left(\frac{\dot{N}_w\sin\phi}{4\pi r^2\sin(\beta+\phi)}
\right)
\end{split}
\end{equation}
in a system where a symmetric, expanding circumstellar envelope is subject to plane-parallel photoionization. $r$ is the distance from the star to the ionization surface, $\beta$ is the angle between the radius vector and the symmetry axis, $\phi$ is the angle between the radius vector and the tangent to the ionization surface, and $z$ is the path length along the ray. The $\sin(\beta+\phi)$ factor represents the incident flux projected on the front \citep{2006AJ....131..603S}. This geometry is shown in Figure \ref{fig:geometry}. In this model, the CSM is assumed to be under a steady wind. It is ambiguous whether the arc was produced by an approximately constant wind as modeled here, or from a distinct ejection episode. Due to the limitations of the optical data, only the former is tested in this work; future spectroscopic velocity measurements can distinguish between the two. Where $\beta=0$, $\sin \phi=1$, the ionization flux equation is simplified to
\begin{equation}
I=\alpha\left(\frac{\dot{N}_w}{4\pi v}
\right)^2 \frac{1}{3r_0^3}+\frac{\dot{N}_w}{4\pi r_0^2}
\label{eq:flux_simplified}
\end{equation}
The steady state mass continuity equation is
\begin{equation}
    \dot{M}\equiv4\pi r^2 \rho v=\text{const}
\end{equation}
where $\rho=\mu_\mathrm{H}m_\mathrm{H}n$. Equivalently, as $\mu_H m_H$ is the mass per hydrogen nucleus, it can be written as $\dot{N}_w\equiv\dot{M}/\mu_H m_H=4\pi r^2 nv$ for the rate at which the star is shedding $H$ nuclei. This is solved for $n$, defined as
\begin{equation}
n(r)=\frac{\dot{M}}{4\pi\mu_{\rm H}m_{\rm H}vr^{2}}
     \equiv n_{0}\left(\frac{r_{0}}{r}\right)^{2}
\label{eq:wind}
\end{equation}
and substituted into Equation \ref{eq:flux_simplified} to obtain
\begin{equation}
I = \frac{\alpha r_{0}}{3}\,n_{0}^{2} + vn_{0}
\label{eq:balance}
\end{equation}
\begin{equation}
\dot{M}=4\pi r_{0}^{2}v\mu_{\rm H}m_{\rm H}n_{0}
\label{eq:mdot}
\end{equation}

The density $n_0$ is estimated as $(\mathrm{EM}/L)^{1/2}$, where $L\approx\pi r_0/2=0.92\,\mathrm{pc}$. Using the measured properties $n_0\approx10.7\,\mathrm{cm^{-3}}$, $r_0=0.59\,\mathrm{pc}$, $\alpha=2.59\times10^{-13}\,\mathrm{cm^3\,s}^{-1}$ (case B recombination), and $v=v_\infty=24.3\,\mathrm{km\,s^{-1}}$ (CO terminal velocity; \citealt{De_Beck_2010}), the ionizing flux is estimated as $I=4.4^{+1.6}_{-1.2}\times10^7\,\mathrm{ph\,cm^{-2}\,s^{-1}}$ (the flux required to sustain the emission measure $\mathrm{EM}=105^{+57}_{-37}\,\mathrm{pc\,cm}^{-6}$). This is exceeded by the total ionizing flux of HD 313822 $\sim6\times10^{8}\,\mathrm{ph\,cm^{-2}\,s^{-1}}$, and may be consistent, considering that the estimated flux is obtained from the lower limit of separation. Using the derived $n_0$, the mass loss from the arc front with $\mu_\mathrm{H} m_\mathrm{H}=2.34\times 10^{-24}\,\mathrm{g}$ is $\dot{M}\approx3.9^{+1.2}_{-0.9}\times10^{-5}\,\mathrm{M_\odot}\,\mathrm{yr^{-1}}$. This value is a lower bound; a potential greater density would result in a larger $\dot{M}$. If a lower ratio [\ion{N}{2}]/H$\alpha$ is assumed $\sim1$, the value would be raised to $\dot{M}\approx5.1^{+1.5}_{-1.2}\times10^{-5}\,\mathrm{M_\odot\,yr}^{-1}$. A shift in $A_V$ to the range $1.0\mathrm{-}3.0$ would propagate as $\dot{M}\approx3.3^{+1.0}_{-0.8}\mathrm{-}6.9^{+2.1}_{-1.6}\times10^{-5}\,\mathrm{M_\odot\,yr}^{-1}$.

It is also important to note that the measurement does not reflect the mass-loss at the time of a potential previous ejection, as the wind speed $v$ is assumed to be steady. A spectrum of the arc will provide an independent measurement of its velocity and the previous mass loss rate. The derived $\dot{M}$ is similar to literature mass loss rate estimates of VX Sgr, e.g. recently $\sim2.0\times10^{-5}\,\mathrm{M_\odot\,yr^{-1}}$ \citep{Gail_2020}, $\sim5\times10^{-5}\,\mathrm{M_\odot\,yr^{-1}}$ \citet{Gordon_2018}, and $\sim1.1\times10^{-5}\,\mathrm{M_\odot\,yr^{-1}}$ \citep{Mauron_Josselin_2011}. 

The values are considerably greater than predicted by the $\dot{M}-$luminosity relation by \citet{10.1093/mnras/staa255} of $\lesssim3\times10^{-6}\,\mathrm{M_\odot\,yr}^{-1}$ for a $\log(L/\mathrm{L}_\odot)\sim5.3$ star with an initial mass $\gtrsim20\,\mathrm{M}_\odot$, which is calibrated on RSGs in their ordinary phases. For the case of VX Sgr, the $\dot{M\mathrm{-}L}$ relation for dust-enshrouded RSGs by \citet{2005A&A...438..273V} is more applicable while it is in an extreme superwind phase (similar to NML Cyg and VY CMa), and is distinct from the mass-loss rates of RSGs outside of this phase (e.g. \citealt{2017MNRAS.465..403G, 2025Galax..13...72V}). The relation by \citet{2005A&A...438..273V} predicts $\sim3.4\times10^{-5}\,\mathrm{M_\odot\,yr}^{-1}$ for VX Sgr and is consistent with the estimate derived from the arc.

\section{Supernova--Circumstellar Nebula Interaction}

The distribution of circumstellar material (CSM) surrounding an RSG/RHG has significant effects on the observable properties of the resulting supernova (SN). These are often spectroscopically classified as Type IIn (e.g. \citealt{1997ARA&A..35..309F}) or Type Ibn (e.g. \citealt{2008MNRAS.389..113P}), displaying narrow emission lines. Numerous studies have analyzed
 the effects of spherical and disk-like CSM interactions with SN ejecta (e.g. \citealt{1982ApJ...258..790C,10.1111/j.1365-2966.2010.16851.x,Harris_2016,McDowell_2018}) in compact CSM shells $r\sim10^{15}\,\mathrm{cm}$.

While small-scale interactions directly influence the SN spectrum in the days following core-collapse, the large-scale ejecta--CSM interaction can shape the resulting supernova remnant over the subsequent hundreds of years. \citet{Orlando_2022} argues that the significant asymmetries in the reverse shock of Cassiopeia A (Cas A; $-4000$ to $+6000\,\mathrm{km\,s^{-1}}$ radial velocity, \citet{Milisavljevic_2013}) are the result of a previous interaction between the SN ejecta and an asymmetric, dense CSM shell. In this model, the SNR propagation velocity is slowed by the CSM material, and a reflected shock is driven inward, colliding with the reverse shock. Using 3D hydrodynamic simulations, \citet{Orlando_2022} finds that the best-fit model for the CSM is a large-scale $r\sim1.5\,\mathrm{pc}$, thin $\sigma\sim0.02\,\mathrm{pc}$ shell with a mass $\sim2\,\mathrm{M_\odot}$. The CSM is proposed to have resulted from an RSG eruption $10^4\,\mathrm{-}10^5\,\mathrm{yr}$ prior to the SN. 

Considering the similarities between the Cas A/large-scale, asymmetric CSM model and the observed arc in this work, it is speculatively explored whether the environment of VX Sgr may result in a similar interaction. \citet{Dwarkadas_2005} finds that the evolution of SNe in circumstellar shells from previous mass-loss ejections depends primarily on a single parameter, $\Lambda$, the ratio of the dense shell mass to the ejected SN material mass. For values $\Lambda\ll1$, the shell will have no significant impact on the SN ejecta, which will quickly destroy the shell. For values $\Lambda \gg 1$, the shell will greatly alter the evolution of the various phases of the SNR formation.

The ionized mass is estimated as $M_\mathrm{ion}= 4\pi r_0^2 t \mu_\mathrm{H} m_\mathrm{H} n_0\approx0.24\,\mathrm{M}_\odot$, where $t$ is the full width half maximum thickness of the arc $\sim0.15\,\mathrm{pc}$. Considering the shell asymmetry, $M_\mathrm{ion}$ is scaled by 0.5 (the approximate ionization hemisphere) to $\sim0.12\,\mathrm{M}_\odot$. A photoionization-confined shell would also trap some of the continued RSG wind and produce a dense neutral gas layer; simulations of the ionized shell surrounding Betelgeuse by \citet{2014Natur.512..282M} find $\sim20\mathrm{-}35\%$ of the RSG wind can be confined from this mechanism. Thus, the neutral gas mass can be approximated as $0.2\mathrm{-}0.35\times0.5\dot{M}r_0/v\approx0.1\,\mathrm{M}_\odot$. Assuming an ejecta mass $\mathrm{\sim15\,\mathrm{M}_\odot}$ \citep{Sukhbold_2016}, also scaled by 0.5 to $7.5\,\mathrm{M}_\odot$ (canceling the $M_\mathrm{ion}$ scale), the mass ratio is estimated as $\Lambda\approx0.02\mathrm{-}0.04$.

Thus, the CSM will not have a significant effect on the forward shock of the SN. Further study with numerical simulations can be performed with spectroscopic observations of the arc.

\section{The RSG Circumstellar Nebula Population}
\label{population}

The $\sim$parsec-scale CSM nebulae surrounding VX Sgr and NML Cyg \citet{1983ApJ...267..179M, 2006AJ....131..603S} may represent an emerging class of extensive photoionized RSG shells. A deeper understanding of these objects would lend valuable insights into the mass-loss histories of RSGs and the formation of asymmetric SNRs through CSM-ejecta interactions.

\begin{figure*}
    \centering
    \includegraphics[width=1\linewidth]{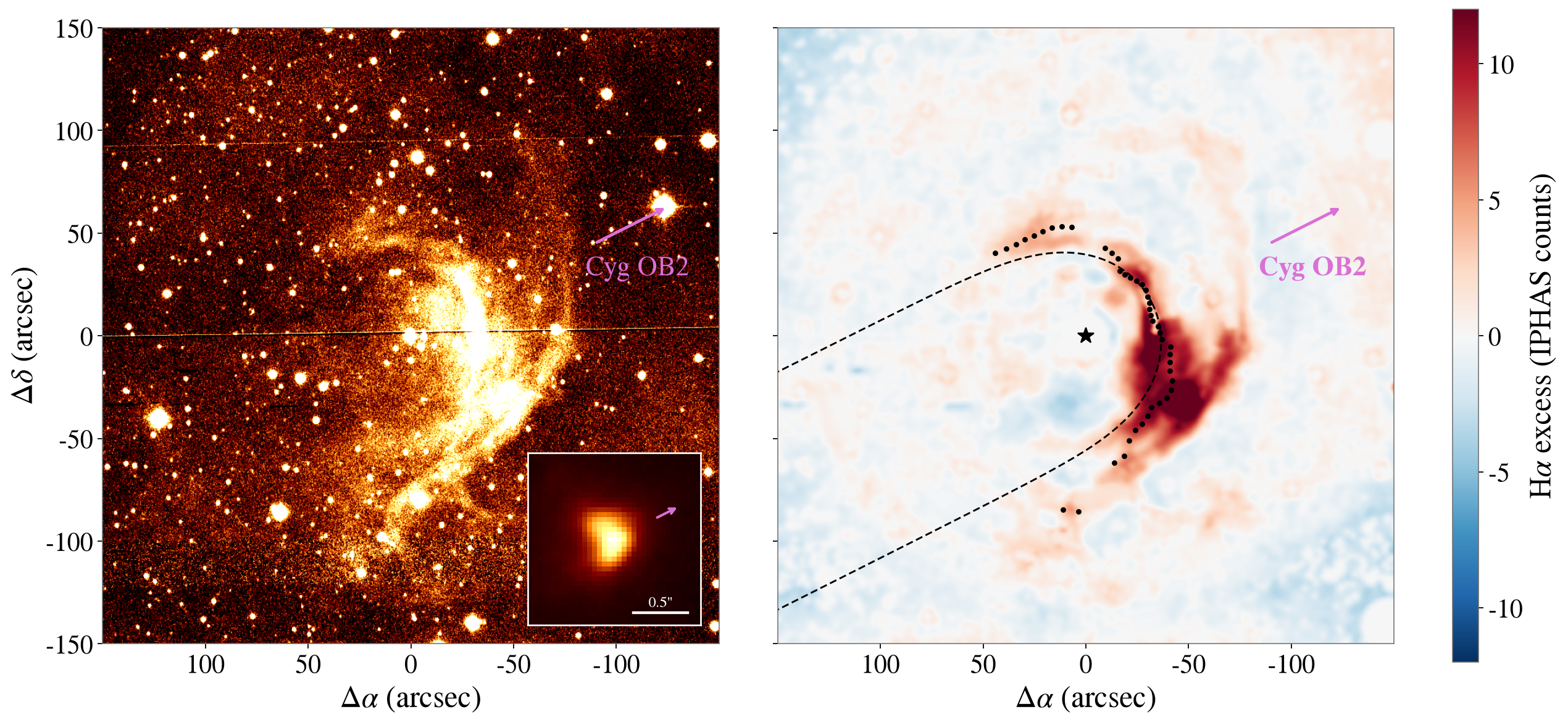}
    \caption{The H$\alpha$ shell of NML Cyg in the IPHAS survey (left) and star-removed excess mask (right). The HST image of the inner envelope described by \citet{2006AJ....131..603S} is shown as an inset in the left panel. Units of the masked image are in IPHAS counts, as the data are not calibrated to R. The pink arrow points towards the potential ionizing source Cyg OB2. N is up and E is left. FOV = $5\times5$ arcmin.}
    \label{fig:nml_cyg}
\end{figure*}

\citet{Zhang_2012} measures the 3D peculiar motion of NML Cyg, finding $\sim8\,\mathrm{km\,s^{-1}}$, which is misaligned with the orientation of the asymmetric CSM shell and is thus not well explained by a bow shock. H$\alpha$ images of NML Cyg are obtained from the INT Photometric Hα Survey of the Northern Galactic Plane (IPHAS; \citealt{2005MNRAS.362..753D, 2014MNRAS.444.3230B}) and are displayed in Figure \ref{fig:nml_cyg}. As the IPHAS data are not calibrated to Rayleighs, the same surface brightness analysis is not performed on NML Cyg in this work. \citet{1982A&A...108..412H} (noted as from H.R. Dickel) reported a H$\alpha$ intensity of $125\pm83\,\mathrm{R}$. Despite the CSM around VX Sgr and NML Cyg both likely being photoionized by hot stars in their respective associations, the two demonstrate prominent differences in their morphologies and emission signatures. 

Compared to the relatively thin shell of VX Sgr, which fits well to a single arc (with the exception of the faint southern region), the CSM of NML Cyg is disturbed and contains multiple shell-like structures. Beyond the inner arc, which is oriented towards Cyg OB2, there is a prominent western ridge. NML Cyg also contains coincident radio emission \citep{1983ApJ...267..179M}. These differences may arise from the local ISM density, mass-loss ejections, and the ionization source (noting that NML Cyg is likely ionized by the Cyg OB2 association generally, rather than an individual source). HST observations by \citet{2006AJ....131..603S} reveal that the far smaller $r_0\sim 170\,\mathrm{AU}$ inner envelope of NML Cyg is also asymmetric and is likely shaped by photodissociation from Cyg OB2. Similarly, the geometric model of \citet{1983ApJ...267..179M} is effectively applied at this smaller scale. Subsequent studies of the dusty ejecta also detect an asymmetric excess of hot dust likely produced by the near-UV radiation flux of Cyg OB2 \citep{2007PhDT........28S, Schuster_2009}. The inner nebula of VX Sgr, also observed by \citet{2006AJ....131..603S}, does not exhibit a similar asymmetry and is instead surrounded by a $\sim50\,\mathrm{AU}$ thick spheroidal envelope.

There are also a handful of large ionized nebulae detected around more distant RSGs. Circumstellar emission is detected $\sim0.1\,\mathrm{pc}$ around a number of RSGs in Westerlund 1 \citep{Fenech_2018}, particularly prominent surrounding W26 \citep{2014MNRAS.437L...1W}. \citet{Siebert_2025} finds a detached outflow $r\sim0.24\,\mathrm{pc}$ surrounding DFK 52 in RSGC2, potentially resulting from a dramatic previous mass-loss event.  

The heating of swept up dust by bow shocks is expected to produce far-infrared emission, which is observed in Herschel $70\,\mu\mathrm{m}$ and $160\,\mu\mathrm{m}$ images for $\mu$ Cep \citep{Cox_2012} and Betelgeuse \citep{1997AJ....114..837N, Decin_2012}. The arc nebulae around VX Sgr and NML Cyg do not have this mechanism and are not observed in any archival far-IR bands.

\begin{figure*}
    \centering
    \includegraphics[width=1\linewidth]{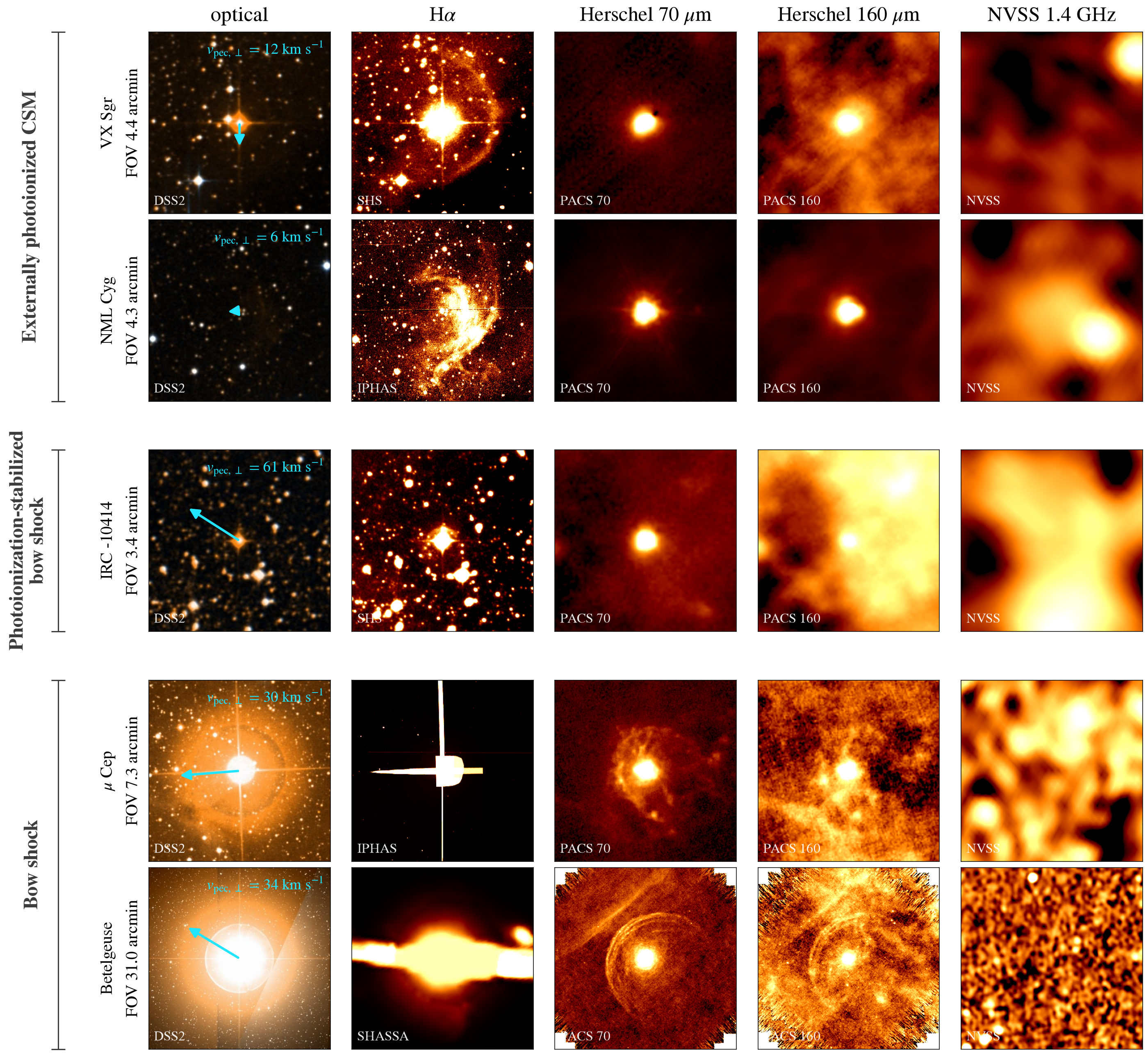}
    \caption{The population of large-scale shells surrounding Milky Way RSG/RHGs. From left to right, the columns show cutout images in optical (DSS2, \citealt{1990AJ.....99.2019L}), H$\alpha$ (SHS, \citealt{2005MNRAS.362..689P, 10.1093/mnras/stt1986}; IPHAS, \citealt{2005MNRAS.362..753D, 2014MNRAS.444.3230B}; SHASSA, \citealt{Gaustad_2001}), 70 $\mu\mathrm{m}$ and 160 $\mu \mathrm{m}$ (Herschel/PACS, \citealt{Pilbratt_2010, Poglitsch_2010}), and $1.4\,\mathrm{GHz}$ (NVSS, \citealt{Condon_1998}), scaled by the angular size of the nebula. Note that the H$\alpha$ images of $\mu$ Cep and Betelgeuse are saturated and contain artifacts, and do not show the bow shocks. Peculiar velocity vectors are shown in cyan, estimated from subtracting Galactic rotation and solar motion
    \citep{2019ApJ...885..131R} from proper motion measurements \citep{2007A&A...474..653V,Zhang_2012,10.1093/mnras/stt1943,Harper_2017,2019ApJ...885..131R}. The objects are arranged by proposed classification as externally photoionized CSM shells, photoionization-stabilized bow shocks, and typical bow shocks. W26 \citep{2014MNRAS.437L...1W} is also a candidate for externally photoionized CSM, but lies too far ($\sim4\,\mathrm{kpc}$) to be resolved in the survey data.}
    \label{fig:rsg_population}
\end{figure*}

IRC --10414 likely represents a case where the conditions of both a bow shock and a photoionization shell are satisfied, as described by \citet{10.1093/mnras/stt1943, 10.1093/mnrasl/slt176}. We are studying the nebula near BI Cyg \citep{1993ApJS...85..137M, 2002ApJ...565.1239L} in Shapiro, J. et al. (in prep.). A comparison of the optical, H$\alpha$, Herschel, and NVSS images of the large scale RSG nebula population is compiled in Figure \ref{fig:rsg_population}.

\section{Conclusion}

In this paper, I describe the discovery of a $\sim$parsec-scale asymmetric shell surrounding the luminous star VX Sagittarii, and characterize the following:\\

1. Thin, arc-shaped emission is detected in H$\alpha$ encompassing VX Sgr with an extinction-corrected intensity $I_{H\alpha}\sim38\,\mathrm{R}$ and radius $r_0\sim0.59\,\mathrm{pc}$. Due to the slow peculiar velocity of VX Sgr and the lack of coincident far-infrared emission in the nebula, it is most likely photoionized circumstellar material resulting from past RSG wind/mass-loss rather than a bow shock. A possible ionization source, the $\mathrm{O}$- or $\mathrm{B}$-type star HD 313822, is explored. It is the most extensive known CSM shell surrounding an RSG.\\

2. Previous studies have debated the nature of VX Sgr as a red supergiant/hypergiant, super-AGB star, or Thorne--Żytkow Object. In this work, adopting membership in the Sgr OB1 association based on modern distance measurements, models of the evolution of RSGs, super-AGBs, and TŻOs are used to constrain stellar models for VX Sgr. It is determined that VX Sgr is not compatible with the evolutionary track of super-AGBs, and is rather likely an RSG/RHG. A spectroscopic measurement of the CSM is proposed for future work to explore its candidacy as a TŻO, which is not excluded.\\

3. The mass-loss rate is estimated using the geometric properties and the incident ionizing flux, expanded from \citet{1983ApJ...267..179M}. The resulting measurement is $\dot{M}\approx4\times10^{-5}\,\mathrm{M_\odot}\,\mathrm{yr^{-1}}$, similar to recent estimates of the mass-loss rate from the dust SED, and consistent with the $\dot{M}$-luminosity relation by \citet{2005A&A...438..273V} applicable to extreme RSGs. Spectroscopic observations of the arc velocity will provide a unique, multi-epoch look at mass-loss around an RSG from distinct periods.\\

4. The interaction between the future SN ejecta of VX Sgr and its $\sim0.1\,\mathrm{M}_\odot$ ionized circumstellar arc is briefly examined with the ratio of shell to SN ejecta mass. The effect of the CSM on the SN shock may not be significant. Spectroscopic follow-up and hydrodynamical simulations will provide a more precise understanding of the potential interaction. \\

5. The large-scale CSM surrounding VX Sgr and NML Cyg are proposed to be of the same class of externally photoionized RSG ejecta shells. Deeper studies of this population may provide important insights into their past mass-loss episodes and future SN-CSM interactions.

\begin{acknowledgments}

I thank the reviewer for their helpful and constructive comments.

This paper makes use of data from the SuperCOSMOS H-alpha Survey (SHS) conducted by the Anglo-Australian Observatory.

Based on data products from observations made with ESO Telescopes at the La Silla Paranal Observatory under programme ID 177.D-3023, as part of the VST Photometric Hα Survey of the Southern Galactic Plane and Bulge (VPHAS+, www.vphas.eu).

I am grateful to collaborators in the MDW H-alpha Sky Survey team (https://www.mdwskysurvey.org; https://mdw.astro.columbia.edu) for access to pre-release DR2 data.

This research made use of hips2fits,\footnote{https://alasky.cds.unistra.fr/hips-image-services/hips2fits} a service provided by CDS.  

This research has made use of the SIMBAD database, operated at CDS, Strasbourg, France.

This paper makes use of data obtained as part of the INT Photometric Hα Survey of the Northern Galactic Plane (IPHAS, www.iphas.org) carried out at the Isaac Newton Telescope (INT). The INT is operated on the island of La Palma by the Isaac Newton Group in the Spanish Observatorio del Roque de los Muchachos of the Instituto de Astrofisica de Canarias. All IPHAS data are processed by the Cambridge Astronomical Survey Unit, at the Institute of Astronomy in Cambridge. The bandmerged DR2 catalogue was assembled at the Centre for Astrophysics Research, University of Hertfordshire, supported by STFC grant ST/J001333/1.

Use of the Southern H-Alpha Sky Survey Atlas (SHASSA), which is supported by the National Science Foundation is acknowledged.

This research is based on observations made with the NASA/ESA Hubble Space Telescope obtained from the Space Telescope Science Institute, which is operated by the Association of Universities for Research in Astronomy, Inc., under NASA contract NAS5–26555. These observations are associated with program GO-8099 (PI: R. M. Humphreys).

This work presents results from the European Space Agency (ESA) space mission Gaia. Gaia data are being processed by the Gaia Data Processing and Analysis Consortium (DPAC). Funding for the DPAC is provided by national institutions, in particular the institutions participating in the Gaia MultiLateral Agreement (MLA). The Gaia mission website is https://www.cosmos.esa.int/gaia. The Gaia archive website is https://archives.esac.esa.int/gaia. 

PACS has been developed by a consortium of institutes led by MPE (Germany) and including UVIE (Austria); KU Leuven, CSL, IMEC (Belgium); CEA, LAM (France); MPIA (Germany); INAF-IFSI/OAA/OAP/OAT, LENS, SISSA (Italy); IAC (Spain). This development has been supported by the funding agencies BMVIT (Austria), ESA-PRODEX (Belgium), CEA/CNES (France), DLR (Germany), ASI/INAF (Italy), and CICYT/MCYT (Spain).

The Digitized Sky Survey was produced at the Space Telescope Science Institute under U.S. Government grant NAG W-2166. The images of these surveys are based on photographic data obtained using the Oschin Schmidt Telescope on Palomar Mountain and the UK Schmidt Telescope. The plates were processed into the present compressed digital form with the permission of these institutions.

The National Radio Astronomy Observatory is a facility of the U.S. National Science Foundation operated under cooperative agreement by Associated Universities, Inc. 

The use of large-language models for enhancing code and assisting in formatting figures is acknowledged.

\end{acknowledgments}

\software{Astropy \citep{astropy:2013, astropy:2018, astropy:2022}, NumPy \citep{harris2020array}, SciPy \citep{2020SciPy-NMeth}, Matplotlib \citep{Hunter:2007}, astroquery \citep{Ginsburg_2019}}

\facilities{UKST, VST, ING:Newton, Herschel (PACS), VLA, Gaia, HST}

\bibliographystyle{aasjournalv7}
\bibliography{references}



\end{document}